\documentclass[preprint,12pt]{elsarticle}
\usepackage{amsmath}
\usepackage{graphicx}  
\usepackage{epstopdf}   
\usepackage{lineno,hyperref}
\modulolinenumbers[5]
\usepackage{color}

\journal{Scripta Materialia}

\begin{document}

\begin{frontmatter}

\title{Relationship between heat effects and shear modulus relaxation during structural relaxation of a telluride glass}


\author[VSPU]{G.V. Afonin\corref{cor1}}
\cortext[cor1]{{Corresponding author}} 
\ead{afoningv@gmail.com, Tel/fax:+7-473-255-24-11} 
\author[VSPU]{R.A. Konchakov}
\author[IEM,DUB]{D.A. Chareev}
\author[IEM,DUB]{S.A. Badmaeva}
\author[STANKIN] {R.S. Khmyrov}
\author[MISIS,MSU] {A.N. Vasiliev} 
\author[IFTT]{N.P. Kobelev}
\author[VSPU]{V.A. Khonik}
\address[VSPU] {Department of General Physics, State Pedagogical
University,  Lenin St. 86, Voronezh 394043, Russia}
\address[IEM] {Korzhinskii Institute of Experimental Mineralogy RAS, Akademik Osip’yan Str. 4, Chernogolovka, 142432 Russia}
\address[DUB] {Dubna State University, Universitetskaya Str. 19, Dubna 141980, Russia}
\address[STANKIN] {Moscow State University of Technology "STANKIN", Moscow 127055, Russia}
\address[MISIS] {National University of Science and Technology MISiS, Moscow 119049, Russia}
\address[MSU] {Lomonosov Moscow State University, Moscow 119917, Russia}
\address[IFTT] {Institute of Solid State Physics RAS, Chernogolovka 142432, Russia}

\begin{abstract}

 We  performed parallel measurements of heat effects and shear modulus relaxation for glassy Te$_{75}$Ge$_{15}$Ga$_{10}$ taken as a representative of practically important non-metallic glasses with covalent bonding. It is shown that  the heat effects occurring upon heating are quantitatively linked to  the shear moduli in the glassy and crystalline states and their temperature derivatives as implied by Eq.(1), which was originally derived for metallic glasses. This relationship provides a good description of exo- and endothermal reactions   using shear modulus relaxation data as an input. This is the first application of this approach to a non-metallic glass with directional interatomic bonding. The obtained results suggest that relaxation phenomena are governed by elastic dipoles -- atomic configurations with the symmetry lower than that of surrounding matrix.

\end{abstract}

\begin{keyword}
\texttt{telluride glasses, calorimetry, shear modulus, relaxation, defects}
\end{keyword}

\end{frontmatter}

Telluride glasses represent a unique class of materials for advanced applications in  photonics, optoelectronics, and sensing due to  combination of their physical and chemical properties \cite{El-Mallawany2011,El-Mallawani2018,Rivera2017}, including ultra-wide infrared transmission window up to 15–35 $\mu$m \cite{LeCoqOptMater2017}, enabling a number of important IR-related applications  \cite{ZhangOptMater2010,ShiryaevJNCS2021,KoltashevOptLasTechn2023,NunesOptLett2021}.

Like all non-crystalline materials, telluride glasses are characterized by excess Gibbs free energy, which constitutes the driving force for spontaneous atomic rearrangements, collectively referred to as structural relaxation (SR). The SR causes significant changes in virtually all physical properties (e.g., enthalpy \cite{SvobodaJNCS2015}, elasticity \cite{LiuJAmCerSoc2014}, anelasticity \cite{GaafarBullMaterSci2015}, volume, viscosity, and refractive index, etc. \cite{El-Mallawany2011}), resulting in the degradation/modification of the performance of glass-based instruments and devices. Meanwhile, available information on the SR of telluride glasses is largely limited to crystallization processes, since the reversible glass $\leftrightarrow$ crystal transition traditionally attracts attention for data storage devices. The evolution of telluride glass properties due to the SR within the non-crystalline state remains largely unexplored. Consequently, a fundamental understanding of the physics governing SR processes is critically required to predict their long-term temporal stability.

To describe relaxation phenomena in glasses, the three-parameter Tool-Narayanaswami-Moynihan (TNM) model is used \cite{Tool1946,Narayanaswamy1971,MoynihanJAmCerSoc1976}. This model operates with the most general characteristics of relaxation processes (non-exponentiality, nonlinearity, and a spectrum of activation energies) and is applied to describe various relaxations irrespective of their physical nature. The TNM model has been successfully used for different types of glasses, including Te-based glasses \cite{SvobodaJNCS2015}. While several other phenomenological models are known \cite{SchererJNCS1990}, the TNM model remains the most widely adopted \cite{Varshneya2019}.

At the same time, developing a modeling framework that can unify the description of relaxations across different physical properties of glasses is of considerable interest. The interstitialcy theory (IT), originally developed and extensively tested for metallic glasses (MGs), offers a promising approach to achieve this. This theory not only provides a rather precise description of various relaxation phenomena in MGs (see  recent review \cite{KobelevUFN2023}), but also  predicts a clear interconnection between physical phenomena that might initially seen as  entirely unrelated. Specifically, the IT establishes a link between thermal effects and  shear modulus relaxation. It is to be emphasized that  the IT was formulated without any assumptions specific to metallic bonding.

The IT assumes the presence of a specific type of defects – split (dumbbell)  interstitials and   describes the properties of glasses using two key equations \cite{GranatoPRL1992,GranatoEurJPhys2014}. The strong dependence of the unrelaxed shear modulus $G$ on the defect concentration $c$ is given as $G = \mu \exp(-\alpha \beta c)$, where $\mu$ is the shear modulus of the counterpart crystal, $\beta$ is the dimensionless shear susceptibility, which constitutes a universal integral parameter relating the shear softening, heat effects, anharmonicity of interatomic interaction and defect structure of glass \cite{MakarovIntermetallics2017}. On the other hand, the shear susceptibility equals the ratio of the 4th-rank (anharmonic) shear modulus to the 2nd-rank (i.e. "usual") shear modulus \cite{MakarovIntermetallics2017,KonchakovJALCOM2017}.  The dimensionless parameter $\alpha \approx 1$ is related to the defect strain field \cite{KobelevUFN2023}. 

The second key assumption relates changes in the internal energy $dU$ to changes in the defect concentration as $\rho dU = \alpha G dc$, where $\rho$ is the density. According to these equations, a change in the defect concentration $c$ leads to a change in the shear modulus $G$, while the latter provides a change in the internal energy $dU$, which, in turn, approximately equals the change in the enthalpy $dH$ and thus gives rise to the  heat effects. This constitutes the physical picture of the heat effects occurring during glass heating. Quantitatively, this picture is described by an equation for the differential heat flow $\Delta W$ \cite{KobelevUFN2023}:

\begin{equation}
\Delta W(T) = \frac{\dot{T}}{\beta\rho}\left[\frac{G(T)}{\mu(T)}\frac{d\mu(T)}{dT} -\frac{dG(T)}{dT}\right],
\label{eq:heatflow_IT}
\end{equation}
where $\dot{T}$ is the heating rate. Given the temperature dependences  $G(T)$ and $\mu(T)$, one  can calculate the heat flow and compare it with experimental calorimetric data.

Meanwhile, interstitial  defects considered by the IT represent a particular case of a broad class of defects – elastic dipoles \cite{Nowick1972}.  The main signature of an elastic dipole is that its symmetry should be lower than that of the surrounding matrix. An elastic dipole can be formed through a wide variety of atomic configurations and, in general, is not related to any specific type of interatomic interaction. This provides expectation that the defects responsible for  formation of non-crystalline structures in various materials may exhibit the properties of elastic dipoles, and consequently, follow the regularities derived within the IT framework. In particular, Eq.(\ref{eq:heatflow_IT})  can be derived by considering the defects in  glass as essentially elastic dipoles, without any reference to dumbbell interstitials \cite{KobelevJApplPhys2014}.

On the other hand, it was earlier shown that the assumption on the presence of elastic dipoles in  a glassy structure leads to exactly the same interrelation between thermal and elastic properties as those in the IT \cite{KobelevJNCS2015,KobelevJApplPhys2014}.   This naturally raises the question of its validity for glasses with different bonding types, such as covalently bonded Te-based glasses—a question the present study aims to address. For this purpose, we chose Te$_{75}$Ge$_{15}$Ga$_{10}$ glass as a representative of a non-metallic, covalently bonded, and practically important telluride glass system.  This glass has a wide optical transparency band from 3 $\mu m$ to 18 $\mu m$ and is perspective for the fabrication of transmitting and acousto-optic devices operating in near infrared and middle infrared ranges \cite{ZhangOptMater2010,KhorkinINCS2025}.

The Te$_{75}$Ge$_{15}$Ga$_{10}$ (at.\%) glass was synthesized from the elements (at least 99.99\% pure) in evacuated quartz ampoules. The elements were weighed in stoichiometric proportions with a total mass of 10 g and placed into an  ampoule (110 mm long, with an inner/outer diameter of 8/12 mm). The melt was held at 650°C for three days. The ampoule was then quickly removed from the furnace and quenched horizontally in cold water. This produced long glassy rods. X-ray study (Panalytical Empyrean diffractometer) confirmed the non-crystalline state. 

Differential scanning calorimetry (DSC) was performed using a Hitachi DSC 7020 instrument calibrated according to the temperatures and enthalpies of melting for In, Sn, Pb and Zn. Measurements were performed in high-purity (99.999\%) nitrogen atmosphere.  The heat flow $W_{gl}$ from an as-cast sample (50–70 mg) was recorded during heating to the complete crystallization temperature $T_{cr}$ at a rate of 3 K/min. The sample was then cooled to room temperature at the same rate. The next DSC scan was performed on the same sample to determine the heat flow $W_{cr}$ in the crystalline state. This procedure enabled the calculation of the differential heat flow $\Delta W = W_{gl} - W_{cr}$. It is this differential heat flow that was used for the calculations with Eq.(\ref{eq:heatflow_IT})   given below. 

The elastic moduli of the glass under investigation were determined using resonant ultrasound spectroscopy (RUS) with a setup similar to that described in Ref. \cite{BalakirevRevSciInstrum2019}. For these measurements, a rectangular parallelepiped sample ($3.6 \times 3.7 \times 4.5$ mm, cut with a low-speed diamond saw) was placed on diagonal corners between two piezoelectric transducers, and all resonances in the 200–1100 kHz range were recorded at room temperature. This resonant spectrum was used to calculate the elastic moduli using special software \cite{BalakirevRevSciInstrum2019}. Next, linear heating (3 K/min) in the range 290 K – 600 K was applied, and the frequency $f$ ($\approx 250$ kHz) of the lowest-lying resonance was continuously measured.  Since temperature changes of the density provide a minor effect on $G$ (less than 3\%), this frequency was used to calculate the shear modulus as a function of temperature, i.e. $G(T) = G_0 \left[ \frac{f(T)}{f_0} \right]^2$, where $f_0$ and $G_0$ are the resonant frequency and shear modulus at room temperature, respectively.

\begin{figure}[t]
\begin{center}
\includegraphics[scale=0.4]{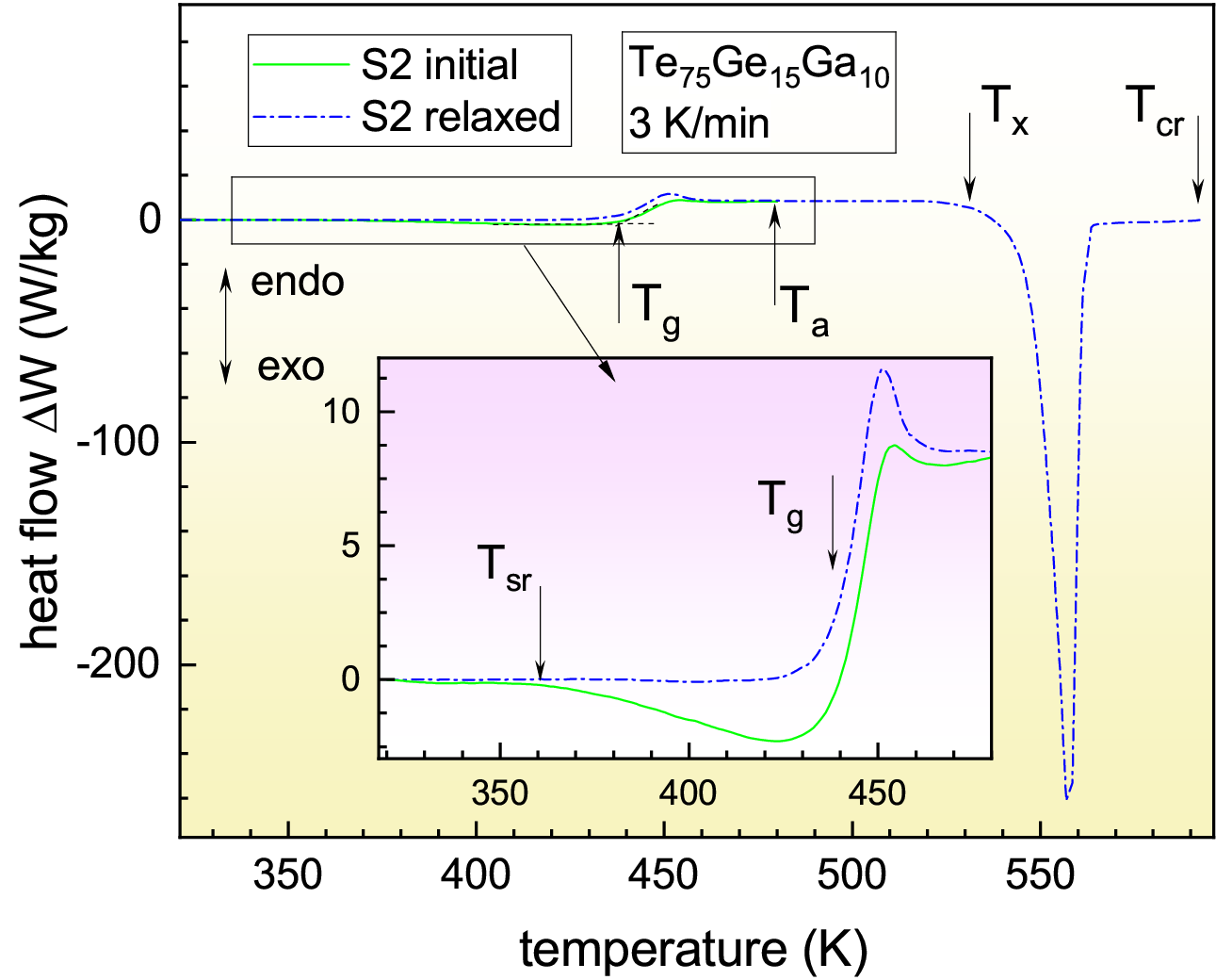}
\caption[*]{\label{Fig1.eps} DSC traces of the same S2  glassy sample in the initial and relaxed states. The characteristic temperatures are indicated: temperature of structural relaxation onset $T_{sr}$,  glass transition temperature $T_g$, preannealing temperature $T_a$, crystallization onset temperature $T_x$, and full crystallization temperature $T_{cr}$.  } 
\end{center}
\end{figure}

RUS and DSC measurements were performed on samples both in the initial and relaxed states. The latter state was obtained by heating a sample to $T_a=468$ K (i.e., into the supercooled liquid state, $\approx 30$ K above $T_g$) and subsequent cooling to room temperature. Full crystallization was achieved by heating to $T_{cr}=600$ K. The RUS spectrum, all elastic constants (longitudinal sound velocity, shear sound velocity, shear modulus, Young's modulus, bulk modulus, Poisson's ratio), and the density at room temperature in the initial, relaxed, and crystallized states are provided in  Supporting Information.

Figure \ref{Fig1.eps} shows the DSC thermogram for the initial and relaxed states. Structural relaxation of the initial sample starts at a temperature $T_a\approx 360$ K, resulting in  exothermal heat flow that changes to  endothermal heat flow at the glass transition temperature $T_g=438$ K. Prior annealing at $T_a$ leads to the complete disappearance of the exothermal relaxation below $T_g$. Overall, the DSC traces presented in Fig.\ref{Fig1.eps} are quite similar to those for MGs, despite  completely different interatomic bonding.

\begin{figure}[t]
\begin{center}
\includegraphics[scale=0.4]{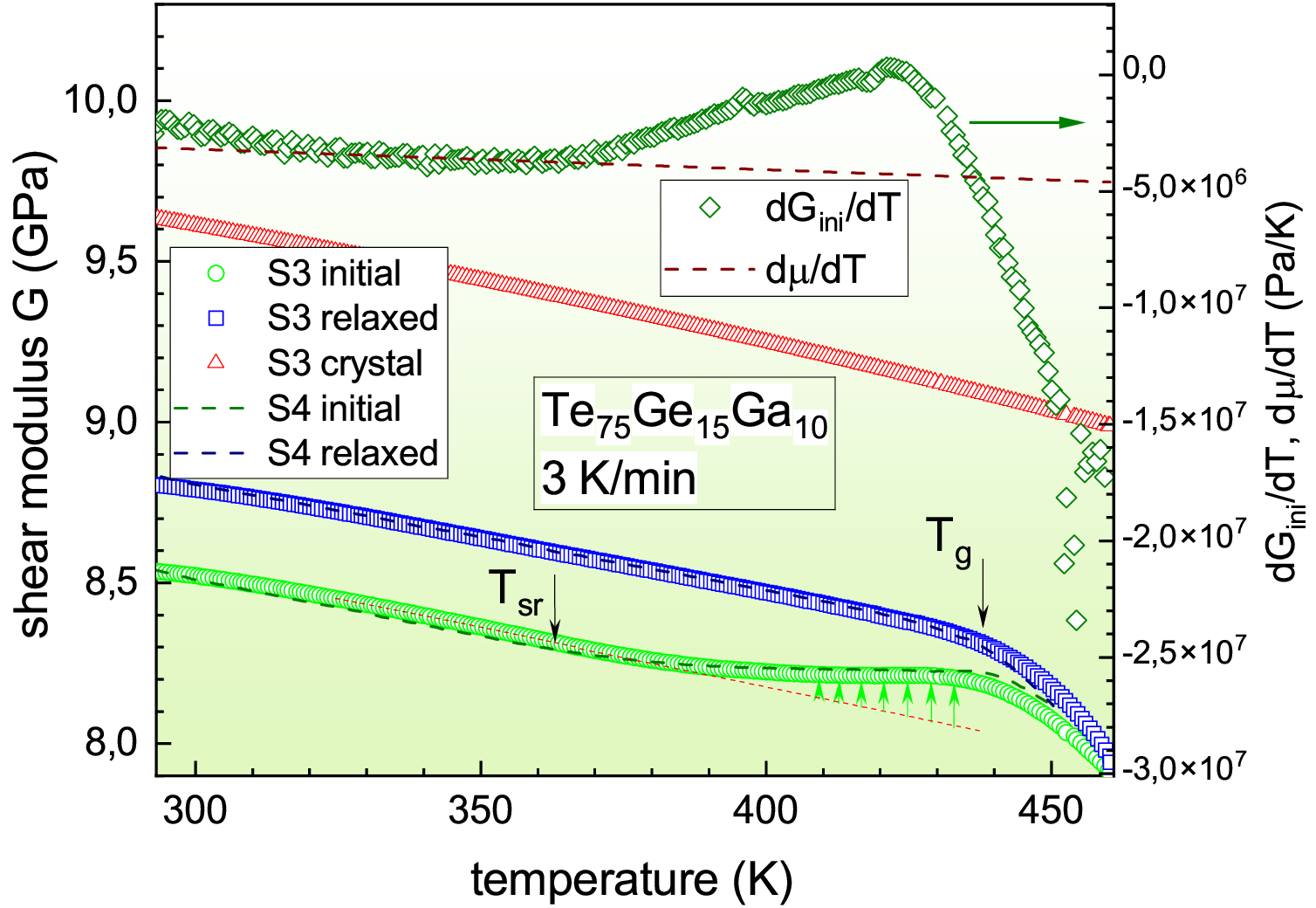}
\caption{Temperature dependence of the shear modulus $G$ for samples S3 and S4 in the initial and relaxed states (symbols and dashed lines), as well as after complete crystallization. The temperature derivatives of the shear modulus in the initial and crystallized states are also presented. 
Temperature $T_{sr}$ marks the onset of structural relaxation, with the upward arrows indicating the progressive relaxation-induced increase of the modulus up to the calorimetric glass transition temperature $T_g$. The large scatter in $dG_{\text{ini}}/dT$ at $T > 450$~K is attributed to deformation of the sample under its own weight above  $T_g$.}
\label{Fig2.eps}
\end{center}
\end{figure} 

Figure \ref{Fig2.eps} presents  temperature dependence of the shear modulus $G$ in the same three structural states. Upon heating of the initial sample, $G$ first decreases linearly due to anharmonicity up to a temperature $T_{sr}\approx 360$ K, which marks the onset of the SR. At higher temperatures, one can observe a relaxation-induced increase in $G$ (indicated by vertical arrows) along with the anharmonic decrease of the shear modulus. Near and above the calorimetric $T_g$, a progressive decrease in $G$ is observed due to entry into the supercooled liquid state and the corresponding drop in viscosity. 

After heating to $T_a=468$ K, the room-temperature $G$ increases by 2.8\% (relaxed state) and then decreases moderately with temperature upon further heating up to $T_g$. At $T>T_g$, one observes softening that is quite similar to that of the initial sample. Continued heating to $T_{cr}=592$ K results in full crystallization (see Fig.\ref{Fig1.eps}). The room-temperature shear modulus increases by $\approx 9.7\%$ and then decreases monotonically with temperature upon subsequent heating. Overall, the shear modulus data in Fig.\ref{Fig2.eps} are fairly similar to the behavior of MGs. It should also be noted that identical measurements carried out on a different sample give essentially the same result (dashed curves in Fig.\ref{Fig2.eps}).

Since Eq.(\ref{eq:heatflow_IT}) contains, in addition to the moduli $G$ and $\mu$ themselves, their temperature derivatives, it is of interest to plot these derivatives separately, as done in Fig.\ref{Fig2.eps}.  It is seen that in the crystalline state, the derivative $d\mu/dT$ decreases smoothly with temperature, whereas in the initial state, the derivative $dG/dT$ increases slightly during structural relaxation below  $T_g$, but then drops rapidly upon transition to the supercooled liquid state.  It is also seen that at about  $T=450$ K,    a strong scatter of points sets in,   obviously caused by the onset of sample deformation under its own weight.  Thus, the use of Eq.(\ref{eq:heatflow_IT}) is limited by this temperature.

We now have all the input data required to calculate the heat flow $\Delta W$ on the basis of shear modulus data (Fig.\ref{Fig2.eps}) using Eq.(\ref{eq:heatflow_IT}). The only quantity in this equation that can be varied to some extent is the shear susceptibility $\beta$.  Within the IT framework, an estimate of $\beta$ can be obtained using the formula $\beta=\frac{G_{rel}-G_{ini}}{\rho \Delta H}$ \cite{KonchakovJALCOM2017}, where $G_{rel}-G_{ini}$ is the change of the shear modulus at room temperature ($T_{RT}$) due to SR, which provides the enthalpy change $\Delta H=\frac{1}{\dot{T}} \int_{T_{RT}}^{T_{cr}} (W_{rel}-W_{ini}) dT$  with $W_{rel}$ and $W_{ini}$ being the heat flows coming from relaxed and initial samples, respectively. This estimate gives $\beta\approx 15$. On the other hand, for different MGs, this quantity varies in a narrow range, $15\leq \beta \leq 25$ \cite{MakarovIntermetallics2017}. Therefore, we accepted $\beta$ as a fitting parameter and found that the same $\beta=15$ gives the best fit. 

\begin{figure}[t]
\begin{center}
\includegraphics[scale=0.85]{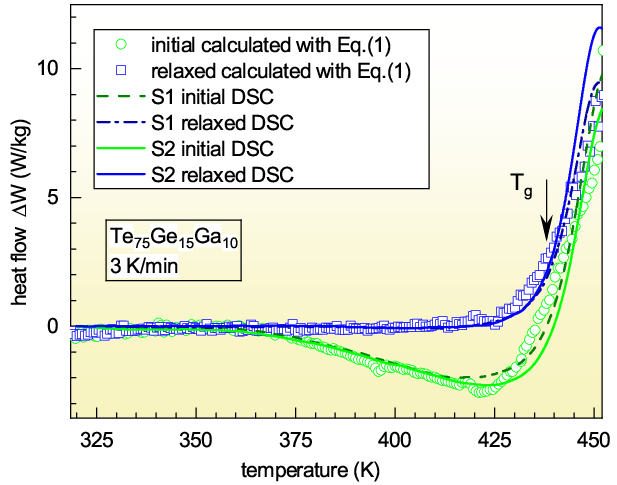}
\caption{Experimental DSC runs on samples S1 and S2 in the initial and relaxed states together with the calculation  of the heat flow using Eq.(\ref{eq:heatflow_IT})  It is seen that this equation provides a good description of exothermal and endothermal relaxation using shear modulus data as an input. }  
\label{Fig3.eps}
\end{center}
\end{figure}

Thus, with the density $\rho = 5.36$~g/cm$^3$ we arrive at the results shown in Fig.\ref{Fig3.eps}. This figure gives the experimental heat flow $\Delta W$ for the samples S1 and S2 in the initial and relaxed states in the  range 320 K$\leq T\leq 453$ K. Second, it gives $\Delta W(T)$-dependence calculated using Eq.(\ref{eq:heatflow_IT}) with the aforementioned $\beta$, $\rho$, $\dot{T}$ and temperature dependences $G(T)$ and $\mu(T)$ shown in Fig.\ref{Fig2.eps}. It is seen that the calculation reproduces the experimental heat flow quite well. For the initial state, Eq.(\ref{eq:heatflow_IT}) replicates the SR-induced exothermal heat flow, which changes into an endothermal reaction near and above  $T_g$. After prior relaxation, the exothermal relaxation disappears completely, in line with the experiment. Therefore, one can conclude that the relationship (\ref{eq:heatflow_IT}), originally derived and tested for MGs, also holds  for a non-metallic glass with directional covalent bonding. 

This conclusion suggests the similarity of defects responsible for the interrelation  between heat effects and shear elasticity relaxation during SR in metallic and telluride glasses. Although defects in glasses cannot yet be unambiguously identified, the prevailing viewpoint is that they are associated with the low‑temperature (boson) peak in the heat capacity, which is characteristic of non-crystalline materials in general. In the case of Te–Ge–Sb telluride glasses, the boson peak  is attributed to dynamic defects \cite{MoesgaardJPCM2025}, understood as regions in the glass where atoms vibrate as independent oscillators with frequencies below the Debye frequency. If one considers only the local structure of dynamic defects (also referred to as quasi‑liquid regions or string‑like defects \cite{HuNatPhys2022}), they are difficult to distinguish unambiguously from the rest of the non‑crystalline matrix. The defining signature of a group of atoms belonging to defects is precisely the pronounced low‑frequency modes in their vibrational spectrum.

A very similar picture of boson peak formation occurs in MGs. Within the IT approach, the boson peak is interpreted as resulting from the contribution of low‑frequency (terahertz) vibrational modes of interstitial dumbbell‑type defects to the vibrational entropy of the glass \cite{GranatoPhysica1996}. The height of the boson peak turns out to be directly proportional to the concentration of such defects, and its change upon thermal treatment  correlates with the relaxation of the shear modulus and with the thermal effects occurring upon SR 
\cite{MitrofanovPSSRRL2019,MakarovIntermetallics2022}. It is also known that dumbbell interstitial defects in metallic glasses form string‑like structures \cite{NordlundEurPhysLett2005}. Thus, the concepts of "string‑like" or "dynamic" defects responsible for the boson peak in Te-based glasses find profound parallels in MGs.

Extrapolating the results of Ref.\cite{MoesgaardJPCM2025} to the studied Te$_{75}$Ge$_{15}$Ga$_{10}$ glass, one can propose the following scenario for the nucleation of dynamic string-like defects. The introduction of Ga and Ge into the tellurium matrix increases the rigidity of local bonds and creates regions with both higher and lower coordination. The Ga and Ge atoms tend toward a four‑coordinated (tetrahedral) environment, which is in conflict with the two‑coordinated chain structure of tellurium. These mismatch regions (frustrated bonds) may act as centers for dynamic defects. 

Thus, the defects in Te‑based glasses can be not just static broken bond, but rather mobile, collective clusters of several tens of atoms that exhibit low‑frequency features in their vibrational spectrum. Exactly the same can be said about defects in MGs, which determines the similarity of relaxation phenomena in these seemingly dissimilar materials. In both cases, the defective regions can be considered as elastic dipoles, which accounts both for the diaelastic effect itself (reduced elastic modulus of  glass) and for the coupling of heat effects with shear elasticity relaxation.

In conclusion, we performed a calorimetric study and measurements of the high-frequency shear modulus on glassy Te$_{75}$Ge$_{15}$Ga$_{10}$, taken as a representative of non-metallic telluride glass with covalent bonding. In an attempt to relate the heat effects and the shear modulus relaxation occurring upon heating, we applied Eq.(\ref{eq:heatflow_IT}), which was developed earlier  for metallic glasses and connects the heat flow and shear modulus relaxation. We show that this equation works  well for both the initial and relaxed states of the glass under investigation. This fact assumes  that the underlying defects -- elastic dipoles -- should exist in a telluride glass and are responsible for the diaelastic effect, structural relaxation, and related heat phenomena.


\begin{thebibliography}{99}

\bibitem{El-Mallawany2011} R. El-Mallawany, Tellurite Glasses Handbook: Physical Properties and Data, Taylor \& Francis, 2011.
\bibitem{El-Mallawani2018} R. El-Mallawany, Tellurite glass smart materials — Application in optics and beyond, Springer, 2018.
\bibitem{Rivera2017} V.A.G. Rivera, D. Manzani (Eds), Technological advances in tellurite glasses, Springer, 2017. 
\bibitem{LeCoqOptMater2017} D. Le Coq, S. Cui, C. Boussard-Plédel, P. Masselin, E. Bychkov, B. Bureau, Telluride glasses with far-infrared transmission up to 35 $\mu$m, Opt. Mater. 72 (2017) 809–812.

\bibitem{ZhangOptMater2010} S. Zhang, X.-h. Zhang, M. Barillot, L. Calvez, C. Boussard, B. Bureau, J. Lucas, V. Kirschner, G. Parent, Purification of Te$_{75}$Ga$_{10}$Ge$_{15}$ glass for far infrared transmitting optics for space application, Optical Materials 32 (2010) 1055–1059.
\bibitem{ShiryaevJNCS2021} V.S. Shiryaev, M.V. Sukhanov, A.P. Velmuzhov, E.V. Karaksina, T.V. Kotereva, G.E. Snopatin, B.I. Denker, B.I. Galagan, S.E. Sverchkov, V.V. Koltashev, V.G. Plotnichenko, Core-clad terbium doped chalcogenide glass fiber with laser action at 5.38 $\mu$m, J. Non-Cryst. Sol. 567 (2021) 120939.  
\bibitem{KoltashevOptLasTechn2023} V.V. Koltashev, B.I. Denker, B.I. Galagan, G.E. Snopatin, M.V. Sukhanov, S.E. Sverchkov, A.P. Velmuzhov, V.G. Plotnichenko, 150 mW Tb$^{3+}$ doped chalcogenide glass fiber laser emitting at $\lambda > 5$ $\mu$m, Optics \& Laser Technol. 161 (2023) 109233. 
\bibitem{NunesOptLett2021} J.J. Nunes, Ł. Sojka, R. W. Crane, D. Furniss, Z. Q. Tang, D. Mabwa, B. Xiao, T. M. Benson, M. Farries, N. Kalfagiannis, E. Barney, S. Phang, A. B. Seddon, and S. Sujecki, Room temperature mid-infrared fiber lasing beyond 5 $\mu$m in chalcogenide glass small-core step index fiber, Optics Lett. 46 (2021) 3504–3507.
\bibitem{SvobodaJNCS2015} R. Svoboda, J. M\'{a}lek,  Enthalpy relaxation kinetics of GeTe$_4$ glass, J. Non-Cryst. Sol. 422 (2015) 51–56.
\bibitem{LiuJAmCerSoc2014} W. Liu, H. Ruan, L. Zhang, Revealing structural relaxation of optical glass through the temperature dependence of Young’s modulus, J. Am. Ceram. Soc. 97 (2014) 3475–3482.
\bibitem{GaafarBullMaterSci2015} M.S. Gaafar, Y.A. Azzam, Acoustic relaxation of some lead niobium tellurite glasses, Bull. Mater. Sci. 38(2015) 119–128.
\bibitem{Tool1946}  A.Q. Tool, Relation between inelastic deformability and thermal expansion of glass in its annealing range, J. Am. Ceram. Soc. 29 (1946) 240–253.
\bibitem{Narayanaswamy1971} O. S. Narayanaswamy, A model of structural relaxation in glass, Am. Ceram. Soc. Bull. 54 (1971) 491–498. 
\bibitem{MoynihanJAmCerSoc1976} C.T. Moynihan, A.J. Easteal, M.A. DeBolt, J. Tucker, Dependence of fictive temperature of glass on cooling rate, J. Am. Ceram. Soc. 59 (1976) 12–16.
\bibitem{SchererJNCS1990} G.W. Scherer, Theories of relaxation, J. Non-Cryst. Solids, 123 (1990) 75–89.
\bibitem{Varshneya2019} A.K. Varshneya, J.C. Mauro,  Fundamentals of inorganic glasses, 3rd ed., Elsevier, Oxford, Cambridge, 2019.
\bibitem{KobelevUFN2023} N.P. Kobelev, V.A. Khonik, A novel view of the nature of formation of metallic glasses, their structural relaxation, and crystallization, Physics--Uspekhi 66 (2023) 673-690. 
\bibitem{GranatoPRL1992} A.V. Granato, Interstitialcy model for condensed matter states of face-centered-cubic metals, Phys. Rev. Lett. 68 (1992) 974-977.
\bibitem{GranatoEurJPhys2014} A.V. Granato, Interstitialcy theory of simple condensed matter, Eur. J. Phys. B 87 (2014) 18.
\bibitem{Nowick1972} A.S. Nowick, B.S. Berry, Anelastic Relaxation in Crystalline Solids, Academic Press, New York, London, 1972.
\bibitem{MakarovIntermetallics2017}  A.S. Makarov, Yu.P. Mitrofanov, G.V. Afonin, N.P. Kobelev, V.A. Khonik, Shear susceptibility - a universal integral parameter relating the shear softening, heat effects, anharmonicity of interatomic interaction and "defect" structure of metallic glasses. Intermetallics 87 (2017) 1-5.
\bibitem{KonchakovJALCOM2017}	R.A. Konchakov, A.S. Makarov, G.V. Afonin, Yu.P. Mitrofanov, N.P. Kobelev, V.A. Khonik, Estimate of the fourth-rank shear modulus in metallic glasses, J. Alloys Compd. 714 (2017) 168-171.
\bibitem{KobelevJNCS2015} N.P. Kobelev, V.A. Khonik, Theoretical analysis of the interconnection between the shear elasticity and heat effects in metallic glasses, J. Non-Cryst. Sol. 427 (2015) 184-190.
\bibitem{KobelevJApplPhys2014} N.P. Kobelev, V.A. Khonik, A.S. Makarov, G.V. Afonin, Yu.P. Mitrofanov, On the nature of heat effects and shear modulus softening in metallic glasses: a generalized approach, J. Appl. Phys. 115 (2014) 033513.
\bibitem{ZhangOptMater2010} S. Zhang, X.-h. Zhang, M. Barillot, L. Calvez, C. Boussard, B. Bureau, J. Lucas, V. Kirschner, G. Parent, Purification of Te$_{75}$Ge$_{15}$Ga$_{10}$ glass for far infrared transmitting optics for space application, Optical Materials 32 (2010) 1055-1059.
\bibitem{KhorkinINCS2025} V.S. Khorkin, E.I. Kostyleva, S.N. Mantsevich, A.P. Velmuzhov, E.A. Tyurina, M.V. Sukhanov, V.S. Shiryaev, Acoustic and acousto-optic properties of Ga$_{10}$Ge$_{15}$Te$_{75}$ glass, J. Non-Cryst. Sol. 648 (2025) 123299.
\bibitem{BalakirevRevSciInstrum2019} F.F. Balakirev, S.M. Ennaceur, R. J. Migliori, B. Maiorov, A. Migliori, Resonant ultrasound spectroscopy: the essential toolbox, Rev. Sci. Instrum. 90 (2019) 121401.
\bibitem{MoesgaardJPCM2025} J. Moesgaard, T. Fujita, S. Wei, Unveiling the boson peaks in amorphous phase-change materials,  J. Phys.: Condens. Matter 37 (2025) 025101.
\bibitem{HuNatPhys2022} Y.-C. Hu, H. Tanaka H Origin of the boson peak in amorphous solids, Nature Phys. 18 (2022 ) 669–677.
\bibitem{GranatoPhysica1996} A.V. Granato, Interstitial resonance modes as a source of the boson peak in glasses and liquids, Physica B 219-220 (1996) 270-272.
\bibitem{MitrofanovPSSRRL2019} Yu.P. Mitrofanov, A.S. Makarov, G.V. Afonin, K.V. Zakharov, A.N. Vasiliev, N.P. Kobelev, G. Wilde, V.A.  Khonik. Relationship between the boson heat capacity peak and the excess enthalpy of a metallic glass. Phys. Stat. Sol. RRL 13 (2019) 1900046.
\bibitem{MakarovIntermetallics2022}  A. Makarov, G. Afonin, K. Zakharov, A. Vasiliev, J. Qiao, N. Kobelev, V. Khonik, Boson heat capacity peak and its evolution with the enthalpy state and defect concentration in a high entropy bulk metallic glass, Intermetallics 141, (2022) 107422.
\bibitem{NordlundEurPhysLett2005} K. Nordlund, Y. Ashkenazy, R.S. Averback, A.V. Granato, Strings and interstitials in liquids, glasses and crystals, Europhys. Lett.  71 (2005) 625-631.




\end{thebibliography}
\end{document}